\documentclass[osajnl,preprint,showpacs]{revtex4}
\DeclareRobustCommand{\baselinestretch{2}}

\begin{document}

\title{Fano resonance in two-dimensional optical waveguide arrays with a bi-modal defect}

\author{Rodrigo A. Vicencio}
\author{Andrey V. Gorbach}
\author{Sergej Flach}
\affiliation{Max-Planck-Institut f\"ur Physik komplexer Systeme, N\"othnitzer Strasse 38, D-01187 Dresden, Germany}

\begin{abstract}
We study the two-dimensional extension of the Fano-Anderson model on the basis of a two-dimensional optical waveguide array with a bi-modal defect.
We demonstrate numerically the persistence of the Fano resonance in wavepacket scattering process by the defect. An analytical approximation is derived for the total scattered light power.
\end{abstract}


\maketitle

Optical waveguide arrays are known to be very efficient experimental realizations of one- (1D) and two-dimensional (2D) discrete systems \cite{Yariv}. Recent
achievements both in sample preparation techniques and in experimental methods allowed for observation of several important
phenomena known from linear and nonlinear lattice theory \cite{NWG_rev}. These experimental successes, in turn, stimulated theorists in
considering other possible implementations of waveguide arrays, being of potential interest both in fundamental and applied science. 
One of such
promising directions of research, being extensively developed during recent years, is related to the possibility of direct observation of the Fano resonance 
\cite{Fano} in processes of wave scattering by nonlinear localized excitations \cite{Fano_DB, Fano_opt} and
in the framework of different linear and nonlinear generalizations \cite{FA_gen} of the Fano-Anderson model \cite{Fano, Anderson, fanoander}. 
In all such models a defect (either geometrical or dynamical) is locally attached to an otherwise translationary invariant 1D waveguide array. This defect mediates one or several additional propagation paths for waves, locally increasing the dimensionality of the system. The Fano resonance manifests through the resonant reflection of light (as the result of destructive interference between the propagation paths) at a particular angle \cite{Fano_opt}, defined by the defect parameters. 

\begin{figure}[t]
  \includegraphics{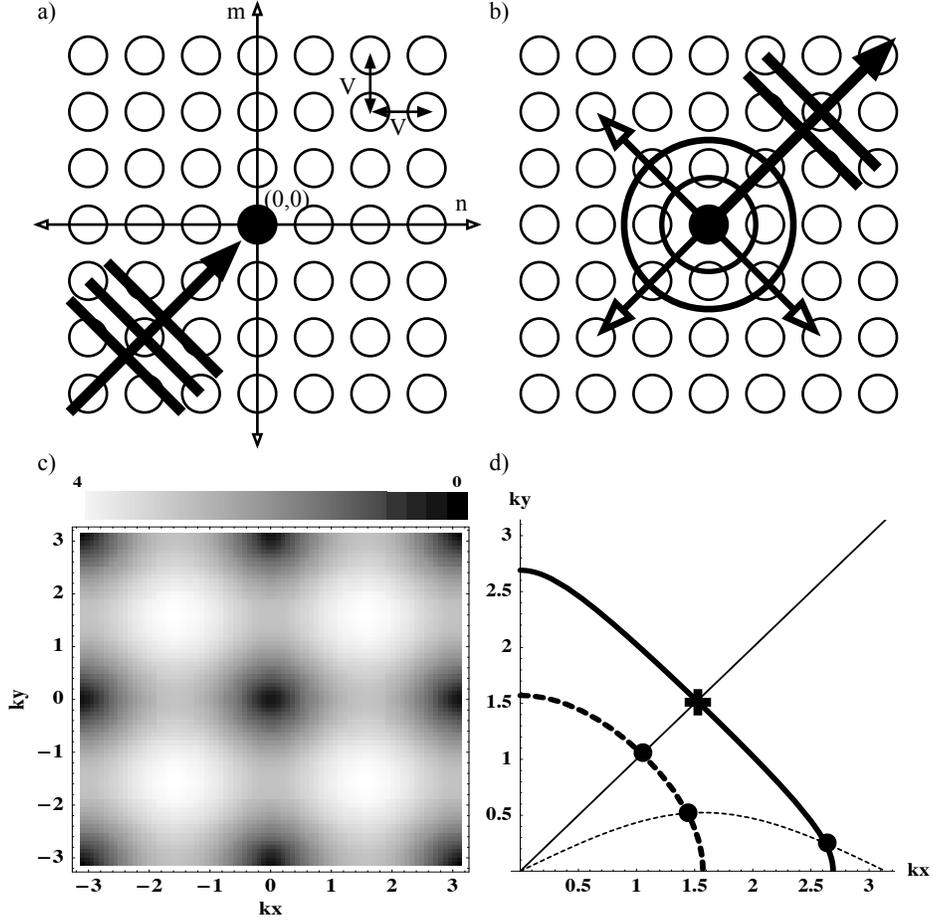}
  \caption{In a) and b) the scheme of a 2D waveguide array with the defect on the central site, and the concept for incoming, reflected, and transmitted waves in the scattering problem are shown; c) Density plot of the plane waves group velocity (absolute value), $V=2$; d) Resonance diagram: Thick solid and dashed lines represent the resonance condition curve given by (\ref{res_cond}) for $E=0.1$ and $E=1$, respectively. Thin solid and dashed lines represent the $(k_x,k_y)$ parameter submanifold corresponding to the \emph{uni-directional} motion of the wavepacket with $v_y=v_x$ and $v_y=v_x/2$, respectively.}
  \label{fig1}
\end{figure}

In this letter we consider the two-dimensional extension of the Fano-Anderson model. The setup is based on a 2D waveguide array, see Fig.~\ref{fig1}(a). An additional defect is assumed to be locally coupled to the central waveguide $\{0,0\}$. Experimentally it can be realized e.g. by making the central waveguide to be bi-modal, so that the defect site corresponds to the second mode of this waveguide. Assuming that the overlap between the two modes of the bi-modal waveguide is much stronger than the overlap between the second mode of the central waveguide and the first (principal) mode of an adjacent waveguide, the coupling to the defect is essentially local. 

Within the framework of coupled-mode theory \cite{Yariv}, the effective model 
equations describing the evolution of the $\{n,m\}$th waveguide mode amplitude $u_{n,m}$ and the defect mode (the second mode of the central bi-modal waveguide) amplitude $\psi$ along the propagation distance $z$ can be written
as \cite{led04}:
\begin{eqnarray}
i \frac{\partial u_{n,m}}{\partial z}+V \Delta u_{n,m}+\epsilon \psi \delta_{n,0} \delta_{m,0}=0\ ,\nonumber\\
i \frac{\partial \psi}{\partial z}+E\psi+\epsilon u_{0,0}=0\ , \label{eq1}
\end{eqnarray}
where $\Delta u_{n,m}\equiv \left (u_{n+1,m}+u_{n-1,m}+u_{n,m+1}+u_{n,m-1}\right)$, $V$ is the effective coupling between modes of the adjacent waveguides,
$\epsilon$ is the effective coupling between the two modes of the central waveguide, and $E$ describes the difference between the propagation constants of the two modes. As usual for Schr\"odinger-type equations, Eqs.~(\ref{eq1}) conserve the total \emph{power} (norm) $P$
\begin{equation}
P=\sum_{n,m} p_{n,m} + p_{def} \equiv \sum_{n,m} |u_{n,m}|^2 + |\psi|^2,
\label{power}
\end{equation}
directly related to the electric field power of a light beam propagating in the array.

Considering stationary solutions $u_{n,m} (z) = A_{n,m} \exp \left[i \omega z\right]$ and $\psi_0=B \exp \left[i \omega z\right]$,
we obtain the following equations for $A_{n,m}$:
\begin{equation}
\omega A_{n,m}=V\Delta A_{n,m}+\frac{\epsilon^2}{(\omega-E)} A_{0,0} \delta_{n,0} \delta_{m,0}.
\label{eq2}
\end{equation}
Far away from the defect site $\{0,0\}$ eigenstates of the system (\ref{eq2}) asymptotically approach plane waves 
$A_{nm}=A \exp \left[i (k_x n +k_y m)\right]$ with the dispersion relation 
and group velocities given by
\begin{eqnarray}
\label{disp}
\omega=2V (\cos k_x +\cos k_y),\\
\label{vg}
\vec{v}_g=-2V (\sin k_x, \sin k_y),
\end{eqnarray}
respectively.
The absolute value of the group velocity (\ref{vg}) is shown in the density plot in Fig.~\ref{fig1}(c). It reaches its maximum at $|k_x|=|k_y|=\pi/2$.

Analogously to the 1D Fano-Anderson model \cite{fanoander}, the defect coupled to the central site acts as an effective scattering potential for traveling waves with strength depending on the spectral parameter ($\omega$) of the incoming wave. The scattering potential diverges at 
$\omega=E$, this constitutes the resonance condition. Combining this with the dispersion relation (\ref{disp}), we obtain the 
implicit relation
\begin{equation}
\cos(k_x)+\cos(k_y)=\frac{E}{2V},
\label{res_cond}
\end{equation}
which defines the \emph{resonance condition curve} in the parameter space $(k_x,k_y)$ plotted in Fig.~\ref{fig1}(c) with thick lines.
Thus, 
for a plane wave
coming at a certain angle (i.e. with the fixed relation between $k_x$ and $k_y$) one has a unique pair of parameter values $(k_x,k_y)$ for which the resonance in the scattering process is expected to occur, see Fig.~\ref{fig1}(d).

We performed numerical simulations of wavepacket scattering in the array of $81\times 81$ waveguides with fixed boundary conditions. The initial wavepacket is taken in the Gaussian form $A_{nm}(z=0)=A \exp \left[i (k_x n +k_y m)\right]$ with $\alpha=0.01$, so that the effective half-width of the 
packet in real space is about $15$ sites, while the half-width in the reciprocal $k$-space is $\Delta_k \approx 0.2$. 
The defect parameter $E$ is chosen to be reasonably small, $E=0.1$, so that the resonance condition curve defined by Eq.~(\ref{res_cond}) in the parameter space $(k_x,k_y)$ passes close enough to the points $|k_x|=|k_y|=\pi/2$ corresponding to the maximum group velocity of the wavepacket. This allows us to save computational time in numerical simulations described below.

In Figs.~\ref{fig2},\ref{fig3} 
characteristic snapshots of the power density $p_{n,m}$ (\ref{power}) distribution across the array, after the scattering process has occurred, are shown  for resonant and non-resonant cases, respectively. While for the non-resonant case the wavepacket remains practically undistorted after passing the defect site, for the resonant case a considerable amount of the initial wavepacket power is scattered in different directions. In order to estimate the amount
of scattered power we define an approximate boundary between ``transmitted'' and ``scattered regions'', see dashed line in Figs.~\ref{fig2},\ref{fig3}. The total scattered power (the power in the "scattered region") as a function of the wavepacket number $k_x$ for the case of diagonal propagation ($k_x=k_y$) is plotted in Fig.~\ref{fig4}. One can clearly observe the resonant peak at $k_x\approx 1.56$, which is in full agreement with the resonance condition $k_x=\arccos[E/(4V)]$ obtained from Eq.~(\ref{res_cond}).

The resonant peak power of scattered light, as well as the characteristic resonance width, depend on the initial wavepacket parameters and the parameters of the defect. In order to have at least a qualitative understanding of the problem, we make a connection to the 1D Fano-Anderson model
by noting that due to the locality of the defect essentially only the diagonal row in the wavepacket interact with the defect site, while the rest passes through practically undistorted, see Fig.~\ref{fig2}. Referring to the exact result for the transmission coefficient of plane waves in 1D case \cite{FA_gen, fanoander}
\begin{equation}
T^{(1D)}(q)=\frac{4V^2 (\omega-E)^2 \sin^2 q }{\epsilon^4+4V^2 (\omega-E)^2 \sin^2 q}
\label{1d}
\end{equation}
($q$ is the wavenumber of the 1D plane wave) and using the Fourier transform of the diagonal part of the initial Gaussian wavepacket, we finally arrive to the following expression for the scattered power:
\begin{equation}
P^{R}(k_x)=\frac{P^{d}}{\sqrt{\pi\alpha}}\int dq \left\{\left[1-T^{(1D)}(q)\right]e^{-(q-k_x)^2/(2\alpha)}\right\},
\label{1d-est}
\end{equation}
where $P^{d}$ is the power of the diagonal row in the initial wavepacket:
\begin{equation}
P^{d}=\sum_{n=m}p_{n,m}(z=0)\approx \sqrt{\frac{\pi}{4\alpha}}.
\label{dpow}
\end{equation}
Despite the simplicity of this 1D approach, it nevertheless gives us a reasonably good estimate for the scattered power, see solid line in Fig.~\ref{fig4}. The noticeable reduction in the peak scattered power is due to the fact, that the effective ratio of the original wavepacket which essentially interacts with the defect is actually more than one row. Analysing the results given by Eqs.~(\ref{1d-est}, \ref{dpow}), we conclude that the Gaussian parameter $\alpha$ performs a twofold influence on the amount of scattered power. First of all, it controls the ratio of the initial wavepacket which effectively interacts with the defect, see Eq.~(\ref{dpow}). The more localized the initial wavepacket is in real space, the higher is the ratio of it which interacts with the defect. On the other hand, the more it is localized in real space, the less it is localized in the reciprocal $k$-space. This, in turn, leads to a broadening of the transmission curve for pure plane waves, making the effective resonance width larger and, at the same time, decreasing the maximum of the scattered power. This broadening effect, however, can be controlled by changing the relative strength of the coupling to the defect $\epsilon/V$. Indeed, the rate of broadening is defined by the ratio of the wavepacket width in $k$-space $\Delta_k$ to the effective resonance width $\Delta_F$, the latter is known for the 1D Fano-Anderson model \cite{Fano,FA_gen, fanoander,Anderson} to be completely determined by the ratio $\epsilon/V$.

To conclude, we discussed the 2D Fano-Anderson model based on a 2D optical waveguide array with a bi-modal central waveguide. Using numerical simulations we demonstrate the persistence of the Fano resonance in this model, which should be detectable experimentally. On the basis of a simple 1D analogue, we derive a reasonably good approximation for the total scattered power of light and study its dependence on the initial wavepacket parameters and the model parameters.

R. A. Vicencio's e-mail address is rodrigov@mpipks-dresden.mpg.de

\newpage

\section*{List of Figure Captions}

Fig. 1. In a) and b) the scheme of a 2D waveguide array with the defect on the central site, and the concept for incoming, reflected, and transmitted waves in the scattering problem are shown; c) Density plot of the plane waves group velocity (absolute value), $V=2$; d) Resonance diagram: Thick solid and dashed lines represent the resonance condition curve given by (\ref{res_cond}) for $E=0.1$ and $E=1$, respectively. Thin solid and dashed lines represent the $(k_x,k_y)$ parameter submanifold corresponding to the \emph{uni-directional} motion of the wavepacket with $v_y=v_x$ and $v_y=v_x/2$, respectively.

Fig. 2. Power density distribution $p_{m,n}$ (\ref{power}) at $z=10$ (after the scattering process). Initial wavepacket parameters are: $k_x=k_y=1.55$ (resonant scattering), $A=0.1$. Coupling constants are: $V=2,\;\epsilon=1$.

Fig. 3. Power density distribution $p_{m,n}$ (\ref{power}) at $z=12.5$ (after the scattering process). Initial wavepacket parameters are: $k_x=k_y=2.2$ (non-resonant scattering), $A=0.1$. Coupling constants are: $V=2,\;\epsilon=1$.

\begin{figure}[t]
\includegraphics{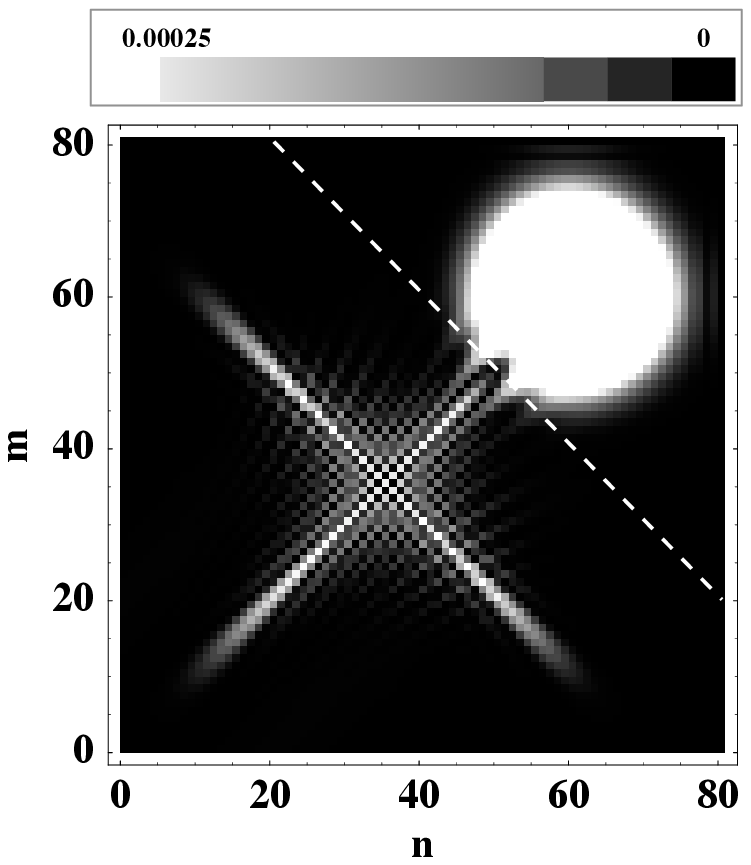}
  \caption{Power density distribution $p_{m,n}$ (\ref{power}) at $z=10$ (after the scattering process). Initial wavepacket parameters are: $k_x=k_y=1.55$ (resonant scattering), $A=0.1$. Coupling constants are: $V=2,\;\epsilon=1$.}
  \label{fig2}
\end{figure}

\begin{figure}[t]
\includegraphics{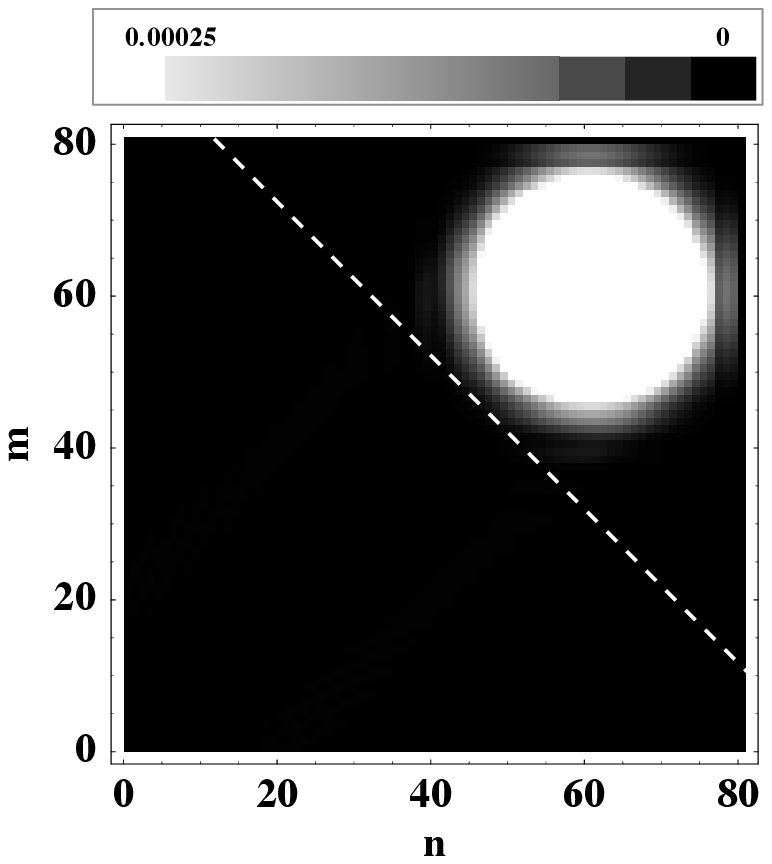}
  \caption{Power density distribution $p_{m,n}$ (\ref{power}) at $z=12.5$ (after the scattering process). Initial wavepacket parameters are: 
  $k_x=k_y=2.2$ (non-resonant scattering), $A=0.1$. Coupling constants are: $V=2,\;\epsilon=1$.}
  \label{fig3}
\end{figure}

\begin{figure}[t]
\includegraphics{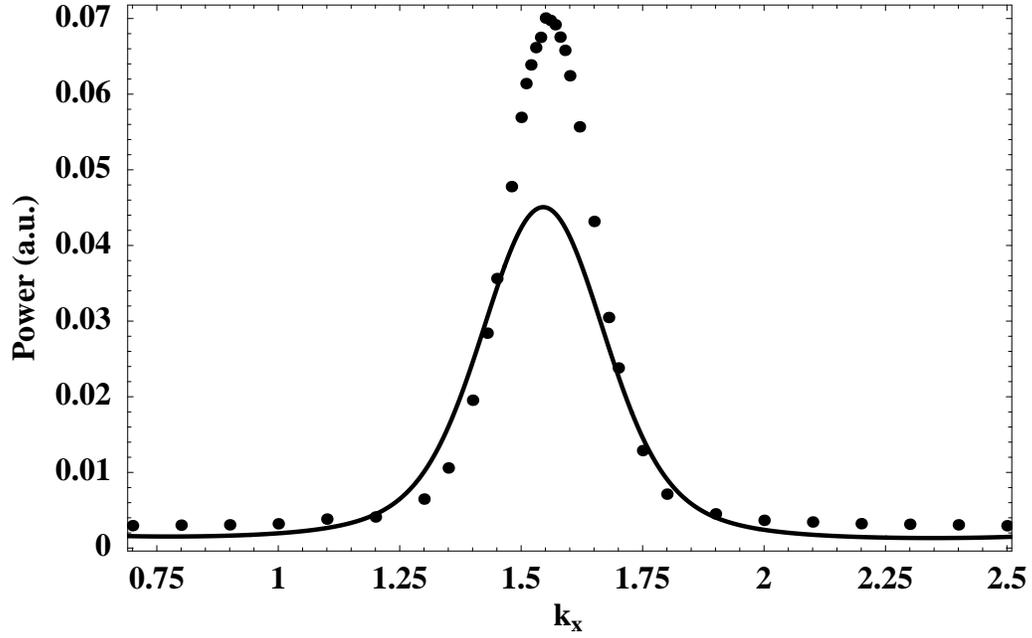}
  \caption{The total scattered power for the diagonal propagating wavepacket ($k_x=k_y$) calculated from numerical simulations (points) and by the approximation (\ref{1d-est}) (continuous line). All the parameter values are the same as in Figs.~\ref{fig2},\ref{fig3}.}
  \label{fig4}
\end{figure}

Fig. 4. The total scattered power for the diagonal propagating wavepacket ($k_x=k_y$) calculated from numerical simulations (points) and by the approximation (\ref{1d-est}) (continuous line). All the parameter values are the same as in Figs.~\ref{fig2},\ref{fig3}.

\newpage

\newpage

\newpage

\newpage

\end{document}